\documentclass{article}
\usepackage{spconf}  

\usepackage{amsmath,epsfig,amssymb,bbold}
\usepackage{bm}
\usepackage{xcolor}
\usepackage{amsthm}
\usepackage[caption=false]{subfig}
\usepackage{cite}
\usepackage[acronym,shortcuts]{glossaries}
\usepackage{lipsum}

\newcommand{\todo}[1]{}
\renewcommand{\todo}[1]{{\color{red} TODO: {#1}}}
\newcommand{\question}[1]{}
\renewcommand{\question}[1]{{\color{red} QUESTION: {#1}}}

\newacronym{GSP}{GSP}{graph signal processing}
\newacronym{GFT}{GFT}{graphical Fourier transform}
\newacronym{DGFT}{DGFT}{digraph GFT}
\newacronym{DV}{DV}{directed variation}
\newacronym{IDV}{IDV}{indefinite directed variation}
\newacronym{CDV}{CDV}{complex directed variation}
\newacronym{TV}{TV}{total variation}
\newacronym{IDGFT}{IDGFT}{indefinite DGFT}
\newacronym{CDGFT}{CDGFT}{complex DGFT}
\newacronym{PCA}{PCA}{principal component analysis}
\newacronym{CS}{CS}{compressive sensing}
\newacronym{FDI}{FDI}{false data injection}
\newacronym{OOBP}{OOBP}{out-of-band power}
\newacronym{LGS}{LGS}{local graph smoothness}
\newacronym{ROC}{ROC}{receiver operating characteristic}

\begin{document} 

\title{Detecting Anomalous Swarming Agents with Graph Signal Processing}
\name{Kevin Schultz\thanks{This work was partially supported by NSF award NCS/FO 1835279 and JHU/APL IR\&D.}, Anshu Saksena, Elizabeth P.~Reilly, Rahul Hingorani, Marisel Villafa\~{n}e-Delgado}
\address{Johns Hopkins University Applied Physics Laboratory\\11100 Johns Hopkins Road, Laurel, MD 20723}




%
\maketitle
\begin{abstract}

 Collective motion among biological organisms such as insects, fish, and birds has motivated considerable interest not only in biology but also in distributed robotic systems. In  a robotic or biological swarm, anomalous agents (whether malfunctioning or nefarious) behave differently than the normal agents and attempt to hide in the ``chaos'' of the swarm. By defining a graph structure between agents in a swarm, we can treat the agents' properties as a graph signal and use tools from the field of graph signal processing to understand local and global swarm properties.  Here, we leverage this idea to show that anomalous agents can be effectively detected using their impacts on the graph Fourier structure of the swarm.
 
\end{abstract}



\section{Introduction}
%

Collective motion in biological systems such as insect swarms, fish schools, and bird flocks are visually striking emergent behaviors that have motivated considerable  research in biology, physics, and engineering \cite{couzin2002collective,sumpter2010collective,vicsek2012collective,o2017oscillators,passino2005biomimicry}.
Due to the distributed nature of swarming systems, graph theory has found considerable utility in the analysis and synthesis of swarming systems by modeling the communications or other interactions between swarming agents as a graph structure
\cite{jadbabaie2003coordination,olfati2006flocking}. Recently, tools from computational topology have been applied to understanding the structure of swarms in terms of connected sub-components as well as the presence of holes and voids 
\cite{topaz2015topological,corcoran2017modelling}.
 These tools explicitly rely on the parametric construction of a graph structure between agents.  

This topological approach was further extended in \cite{schultz2021swarmGSP} to analyze how the local and global ``order'' of swarm properties varies with respect to the graph structure.
The analysis in \cite{schultz2021swarmGSP} employed the field of \ac{GSP} and graph Fourier analysis to show that common swarm states were highly structured (i.e., band-limited) when viewed in the graph Fourier domain. Broadly speaking, \ac{GSP} builds on  its roots in spectral graph theory \cite{chung1997spectral} and algebraic signal processing \cite{puschel2008algebraic}  to generalize concepts from classical signal processing to signals defined on the vertices of irregular domains modeled by graphs \cite{shuman2013emerging,sandryhaila2013discrete,sandryhaila2014discrete,ramakrishna2020user}.

Anomaly detection \cite{chandola2009anomaly} is among the application areas considered in the seminal \ac{GSP} works \cite{sandryhaila2014discrete,egilmez2014spectral}. Since the initial work that considered temperature sensor networks \cite{sandryhaila2014discrete} and more generally abstract sensor networks \cite{egilmez2014spectral}, \ac{GSP}-based anomaly detection has been applied to a number of areas, including power systems \cite{drayer2019detection,ramakrishna2019detection}, social networks 
\cite{ramakrishna2020user}, and image processing 
\cite{verdoja2020graph}.
%
At a high level, these techniques exploit (generally low-pass) structure in the \ac{GFT} of some signal defined on the vertices of a graph, and then threshold on the signal content after (graph) filtering to remove this structure \cite{ramakrishna2020user}.  We note that this class of problem is fundamentally different from detecting an anomalous graph structure in a network, itself a well studied problem \cite{akoglu2015graph}.

In this work, we show how the \ac{GFT} structure of swarms revealed in \cite{schultz2021swarmGSP} can be used to design ``graph filters'' that operate on the swarm state to detect agents within the swarm whose dynamics (and thus behavior) are fundamentally different from the bulk of the swarm. These anomalous agents have behaviors that differ in subtle ways from the rest of the swarm; from a purely kinematic perspective the anomalous trajectories are consistent with the non-anomalous agents.  Instead, these behaviors are detected through interaction and comparison with their neighbors that manifest in the \ac{GFT} domain as outliers.  To our knowledge, this work is the initial adaptation of \ac{GSP} techniques to the swarming domain, and the detection problems herein are challenging enough that multiple swarm measurements are needed for effective detection, demonstrating an anomaly detection problem where the signal is not only time-varying as in \cite{lewenfus2019use}, but with a time varying graph, as well.
This work is related to, but distinct from, the inference of dynamical parameters  \cite{katz2011inferring} and the identification of collective states 
\cite{berger2016classifying,topaz2015topological,corcoran2017modelling}, 
as this work more closely resembles clustering (i.e., unsupervised learning) of distinct behavior regimes within the swarm.  More generally, this work falls under fault detection in swarms \cite{qin2014survey}, and addresses similar problems as \cite{ahn2019learning}, which uses neural networks to identify joint collective anomalies, comprising a more data intensive approach.


In the following, we first review some \ac{GSP} and swarming preliminaries. Next, we briefly discuss how to interpret swarming in a \ac{GSP} context and define two \ac{GSP} approaches to the detection of anomalous agents in a swarm. We then consider both approaches in a range of scenarios using swarming simulations of \cite{couzin2002collective} and \cite{o2017oscillators}, discussing how to leverage their unique graph Fourier signatures, and analyze the resulting anomaly detectors using detection theory. We conclude with summary remarks and discuss future directions.

\section{GSP Background}

A graph $\mathcal{G}=(\mathcal{V},\mathcal{E})$ consists of a collection of $N$ vertices $\mathcal{V}=\{v_i\}_{i=1}^{N} $ and edges $e_{ij} \in \mathcal{E}$ connecting nodes $v_i$ and $v_j$. The graph adjacency matrix $\mathbf{A}\in\mathbb{R}^{N\times N}$ mathematically represents interactions in a graph, with nonzero entries $A_{ij}$ indicating the presence of an edge $e_{ij}$. In this work, we are concerned with non-negative weighted undirected graphs, so $A_{ij}=A_{ji}\geq 0$. The degree matrix $\mathbf{D}$ is a diagonal matrix whose entries account for the total number of connections for each node and is defined as $D_{ii}=\sum_{j}A_{ij}$. The (combinatorial) graph Laplacian is defined as $\mathbf{L}=\mathbf{D}-\mathbf{A}$.
A related matrix, the normalized Laplacian, is defined for connected graphs by $\bar{\mathbf{L}}=\mathbf{D}^{-1/2}\mathbf{L}\mathbf{D}^{-1/2}$ and extends to general graphs several nice properties of the Laplacian that only hold for certain regular graphs \cite{chung1997spectral}. 

\ac{GSP} is concerned with the analysis of signals or functions defined on the vertices of a graph.  Let $\mathbf{f}:\mathcal{V}\to \mathbb{F}^m$ be a so-called graph signal defined on $\mathcal{V}$ that takes values in some finite dimensional Hilbert space.  We adopt the shorthand convention that $\mathbf{f}_i=\mathbf{f}(v_i)$ for the $i$th vertex of $\mathcal{G}$. Over the last decade, most \ac{GSP} efforts have focused on extending techniques defined in classical signal processing to signals defined over graphs, such as filtering and multiple signal transformations including the \ac{GFT}. 
In a natural definition for non-negative weighted symmetric graphs, the \ac{GFT} uses the eigenvectors of $\mathbf{L}$ as basis functions instead of the complex exponentials used in the Fourier transform. However,  \cite{schultz2021swarmGSP} demonstrated that $\bar{\mathbf{L}}$ may be better suited for analysis of swarms. Using the eigendecomposition $\bar{\mathbf{L}}=\mathbf{U}\Lambda \mathbf{U}^\top$, with $\mathbf{U}$ a unitary matrix of eigenvectors, and $\Lambda$ a diagonal matrix of eigenvalues in increasing order,
the \ac{GFT} of a graph signal $\mathbf{f}$ is defined as $\hat{\mathbf{f}}=\mathbf{U}^\top \mathbf{f}$ and the corresponding inverse \ac{GFT} as $\mathbf{f}=\mathbf{U}\hat{\mathbf{f}}$.  Using this definition of the \ac{GFT}, the eigenvalues $\lambda_i$ are a natural generalization of frequency that are no longer evenly spaced in general, but will be in the interval $[0,2]$.  
%
%
%
%
This further suggests an intuitive mechanism to define graph filters using the eigendecomposition. Let $\mathbf{H}$ be a diagonal matrix, then $\mathbf{U}\mathbf{H}\mathbf{U}^\top\mathbf{f}$ is a filtered graph signal where the filter has ``frequency response'' $H_{ii}$ at frequency $\lambda_i$.

\section{Swarm Models}

\begin{figure}[htb]

    \centering
    \includegraphics[width=\columnwidth]{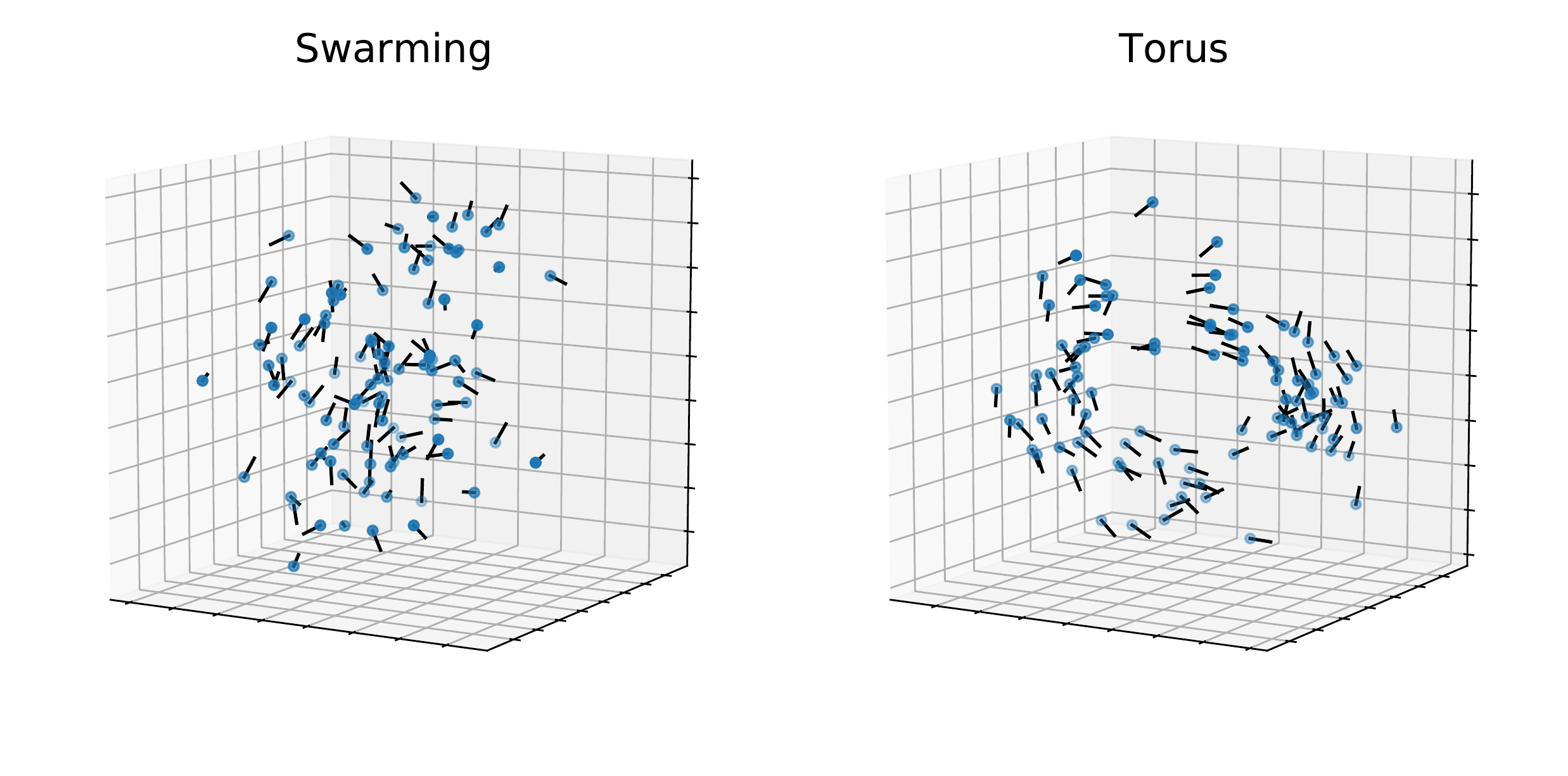}

    \caption{Example swarm states for the model of \cite{couzin2002collective}. Dots indicate agent positions and lines indicate velocities.}

    \label{fig:couzin_baseline_3d}

\end{figure}

In this work, we will consider anomalous agents in an otherwise homogeneous swarm using two different swarming models: the biologically inspired model of \cite{couzin2002collective} and the swarmalator model of \cite{o2017oscillators}.   The model in \cite{couzin2002collective} uses disjoint behavior regions for repulsion, alignment, and attraction.  Depending on the relative radii of these regions, the angle of a blind-spot behind each agent, and the amount of noise, the swarm dynamics will produce one of four steady states. In the absence of an alignment region, the dynamics produce a disorganized ``swarming'' state characterized by low local and global alignment in velocity and minimal collective displacement.
When the region of alignment is larger than the region of repulsion, but still small relative to the region of attraction, the swarm tends to form three-dimensional torus-like structures, with a high level of local alignment in velocity but overall low global alignment and little collective displacement. As the  alignment region approaches the attraction region, two additional collective states form, both with higher global velocity alignment and collective displacement than the swarming and torus states.  Here, we focus on the first two states due to their apparent disorganization, presumably making an anomalous agent harder to detect (see Fig.~\ref{fig:couzin_baseline_3d}).

\begin{figure}[htb]

    \centering
    \includegraphics[width=\columnwidth]{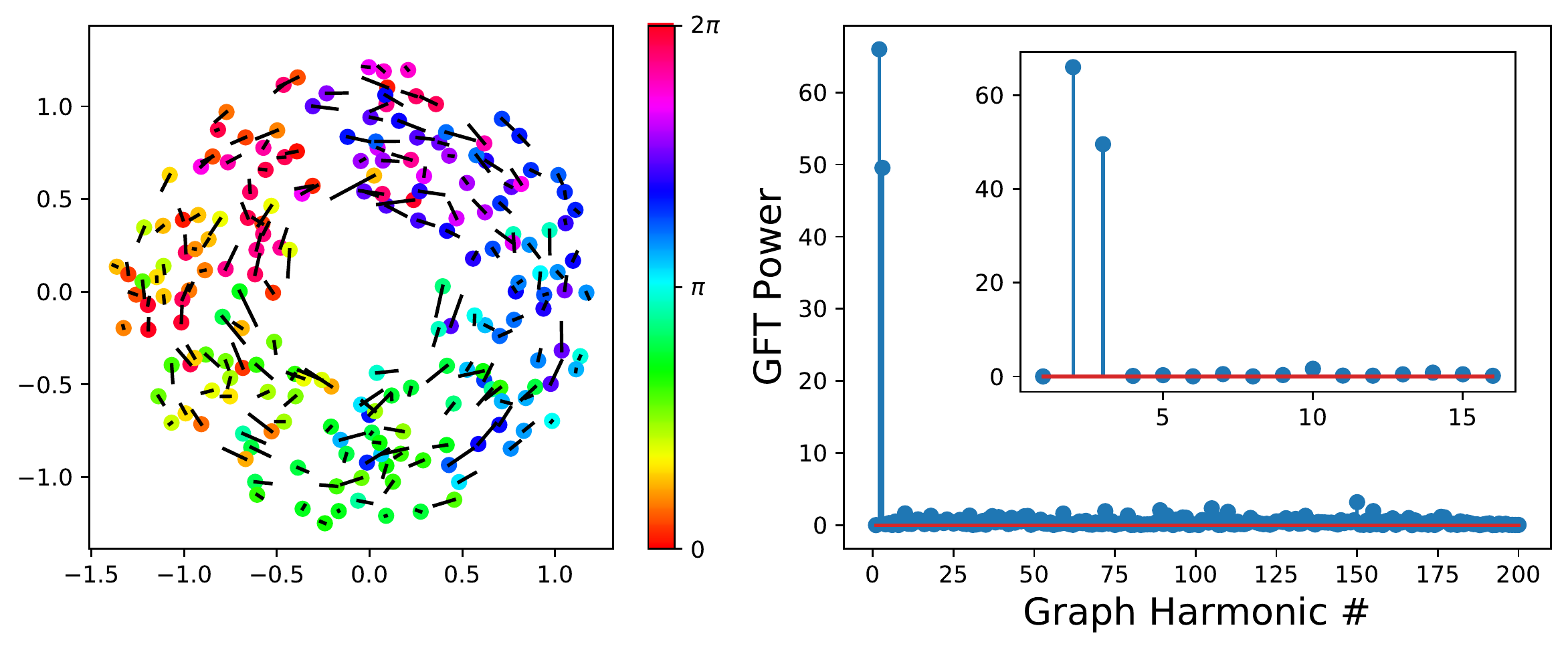}

    \caption{Example swarmalator state. Left: Agent positions, with color indicating $\theta_j$ and lines indicating velocity. Right: \ac{GFT} power using signal $\mathbf{h}$ for the swarm state on the left. Inset axes show zoomed view of spectral concentration.}
    \label{fig:swarmalator_nominal}

\end{figure}

The swarmalator model was developed to understand joint swarming and synchronization behaviors, combining spatial attraction and repulsion with an auxiliary phase $\theta_j$. This phase can modulate the sign of the spatial attraction/repulsion and is itself governed by dynamics in the vein of the Kuramoto model \cite{kuramoto1975international}. 
Of the several steady states observed in \cite{o2017oscillators} we focus on the active wave state in two dimensions, where the agents form counter-rotating (roughly) phase-ordered rings that exhibit considerable dynamism in both the position and phase states (see Fig.~\ref{fig:swarmalator_nominal}, left panel).

\section{Methods}

The necessary ingredients for performing \ac{GSP} analysis are the specification of a graph and a function defined on the vertices of that graph.  For a collection of $N$ swarming entities with positions $\mathbf{x}_j$ and velocities $\mathbf{v}_j$ both $\in \mathbb{R}^n$ (time indices surpressed), following \cite{schultz2021swarmGSP} we define each agent as a vertex in a graph, and for each instance in time we construct an adjacency matrix $A_{jk}=\exp(-||\mathbf{x}_j-\mathbf{x_k}||^2_2/\sigma^2)$ ($j\neq k$, otherwise $A_{jj}=0$), where $\sigma^2=\frac{1}{N(N-1)}\sum_{j\neq k}||\mathbf{x}_j-\mathbf{x}_k||_2^2$. 
Depending on the scenario, we consider different graph signals. For the model of \cite{couzin2002collective} we consider the normalized velocity $\mathbf{u}_j=\mathbf{v}_j/||\mathbf{v}_j||^2_2$ and the adjusted position $\mathbf{r}_j=\mathbf{x}_j-\bar{\mathbf{x}}$ where $\bar{\mathbf{x}}=\frac{1}{N}\sum \mathbf{x}_j$.  For the analysis of swarmalators we use the signal $\mathbf{h}_j=\exp(i\theta_j)$.
%

\begin{figure}[htb]

    \centering
    \includegraphics[width=\columnwidth]{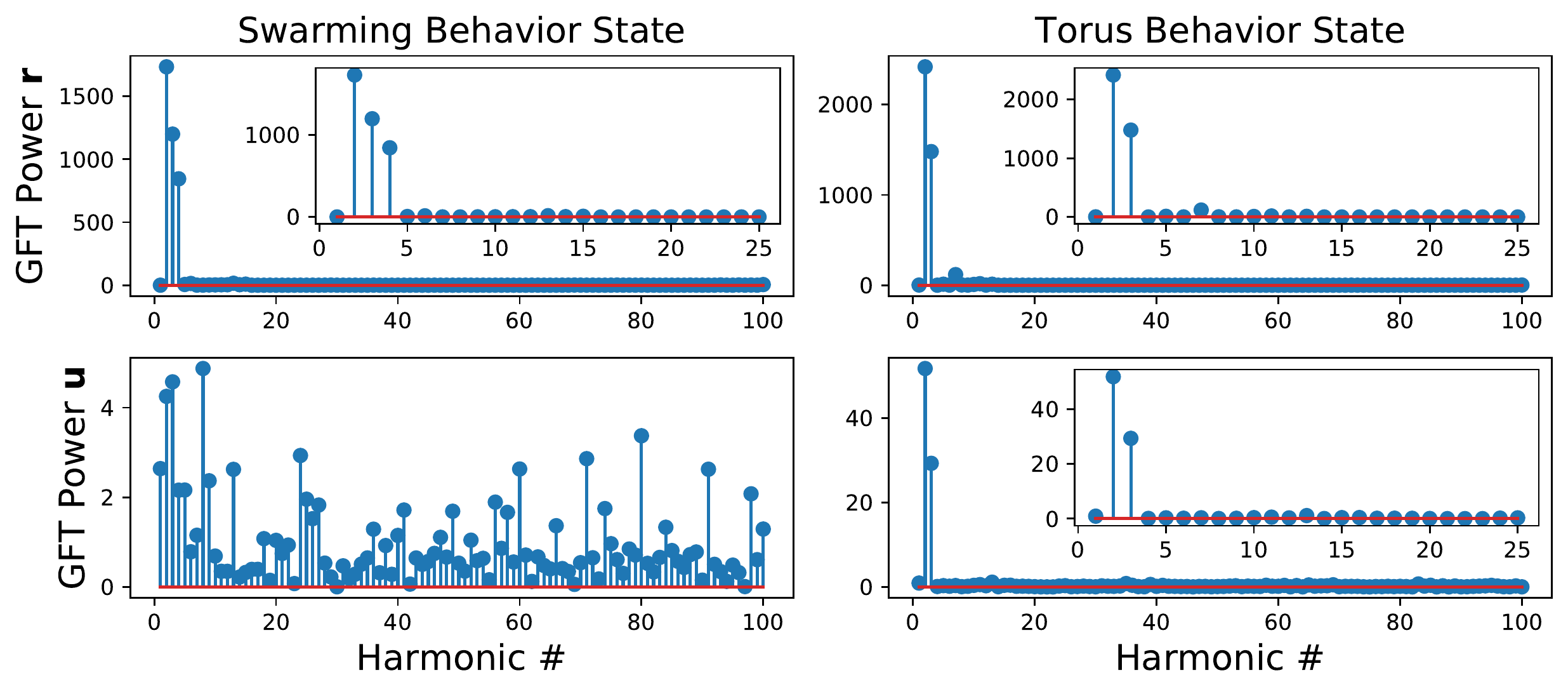}

    \caption{Example \ac{GSP} analysis of swarming (left) and torus (right) states from Fig.~\ref{fig:couzin_baseline_3d}. (Top) \ac{GFT} power of position states $\mathbf{r}$ and (Bottom) \ac{GFT} power of normalized velocity states $\mathbf{u}$.
    Inset axes show zoomed view of spectral concentration.}

    \label{fig:couzin_baseline}

\end{figure}

With these graph and graph signal definitions, using the \ac{GFT} derived from $\bar{\mathbf{L}}$ we have
%
%
from \cite{schultz2021swarmGSP} that both the model of \cite{couzin2002collective} and swarmalators
exhibit spectral concentration in a few graph Fourier harmonics (see Fig.~\ref{fig:couzin_baseline} and Fig.~\ref{fig:swarmalator_nominal}, right panel). Furthermore, these harmonics are low frequency, indicating local alignment of the graph signal with respect to the graph topology.

This spectral concentration immediately suggests two approaches for anomaly detection: 
1) a graph filter approach with filters defined by $H_{ii}=\mathbb{1}_{S}(i)$ where $\mathbb{1}_{S}(i)=1$ if $i\in{S}$ and $0$ otherwise and $S$ is a set of graph frequencies expected not to be common in nominal swarms, and 2) using the intuition that an anomalous agent should be different in ``smoothness'' than the rest of the swarm, the high-pass filter $\mathbf{H}=\Lambda$, which has been suggested as a measure of local graph smoothness for a graph signal \cite{ramakrishna2020user}. We call the first approach \ac{OOBP} and the second \ac{LGS}.  In either case, the input graph signal $\mathbf{f}$ is filtered to generate $\mathbf{g} = \mathbf{U}\mathbf{H}\mathbf{U}^\top \mathbf{f}$, and $||\mathbf{g}_i||^2_2$ is used as a threshold statistic for the detection of an anomalous agent in the swarm (c.f., \cite{egilmez2014spectral,drayer2019detection,ramakrishna2019detection}). 


\section{Results}

We ran the swarming model of \cite{couzin2002collective} with 100 agents (99 normal, 1 anomalous) and the swarmalator model with 200 agents (199 normal, and 1 anomalous).  We ran each with a few different sets of parameters for both the nominal and anomalous agents.  For the model of \cite{couzin2002collective}, we ran 4 cases:
\begin{itemize}
    \setlength{\itemsep}{0pt}
    \setlength{\parskip}{0pt}
    \setlength{\parsep}{0pt}
    \item \textbf{Case 1}: Nominal behavior is Swarming, anomalous agent has slightly larger repulsion region ($\mathbf{r}$ is the graph signal)
    \item \textbf{Case 2}: Nominal behavior is Torus, anomalous agent has slightly larger repulsion region (same anomaly as Case 1, $\mathbf{r}$ is the graph signal)
    \item \textbf{Case 3}: Nominal behavior is Torus, anomalous agent has no alignment region ($\mathbf{u}$ is the graph signal)
    \item \textbf{Case 4}: Nominal behavior is Swarming, anomalous agent has large alignment region ($\mathbf{u}$ is the graph signal) 
\end{itemize}

For the swarmalator model, the nominal behavior of positional attraction to like phases and phase repulsion to nearby phases is defined by parameters $A=B=J=1$, $K=-0.75$ and the anomaly was parameterized at $J=-1$ for positional repulsion to like phases and various values of $K$ in $[-0.75,0.75]$ that varies the strength of the phase repulsion and attraction, using the model definitions as in \cite[Fig.~4]{o2017oscillators}.
We use $\mathbf{h}$ as the graph signal, and we denote this \textbf{Case 5}.

We define \ac{OOBP} filters for each case based on the nominal and anomalous behaviors.  
For both Case 1 and 2, the nominal $\mathbf{r}$ is lowpass (Fig.~\ref{fig:couzin_baseline}, top row) so we use the highpass filters 
$H_{ii}=\mathbb{1}_{\{5,\dots N\}}(i)$ and 
$H_{ii}=\mathbb{1}_{\{4,\dots,N\}}(i)$, respectively. 
For Case 3, $\mathbf{u}$ is lowpass (Fig.~\ref{fig:couzin_baseline}, bottom right) and we use the same filter as Case 2.  
Despite the bandlimited nature of these signals, we found that the exclusion of the first harmonic improved detection.
For Case 4, we expect the nominal swarm state to be spectrally flat (Fig.~\ref{fig:couzin_baseline}, bottom left) and an anomaly with a large alignment range to contribute to low frequency components, so our \ac{OOBP} filter should be lowpass.  We find that $H_{ii}=\mathbb{1}_{\{1,\dots,6\}}(i)$ performed well.  
For Case 5, the signal $\mathbf{h}$ is bandlimited (Fig.~\ref{fig:swarmalator_nominal}, right), and the bandstop filter $H_{ii}=\mathbb{1}_{\{1\}\cup\{4,\dots,N\}}(i)$ produced excellent results.

While the \ac{LGS} approach is already defined by a given swarm's graph, note that since anomalies in Case 4 are expected to appear in low frequencies, they will have low values in the \ac{LGS} approach, but high values for the remaining cases.  We set the direction of our detector accordingly.

\begin{figure}[htb]

    \centering
    \includegraphics[width=\columnwidth]{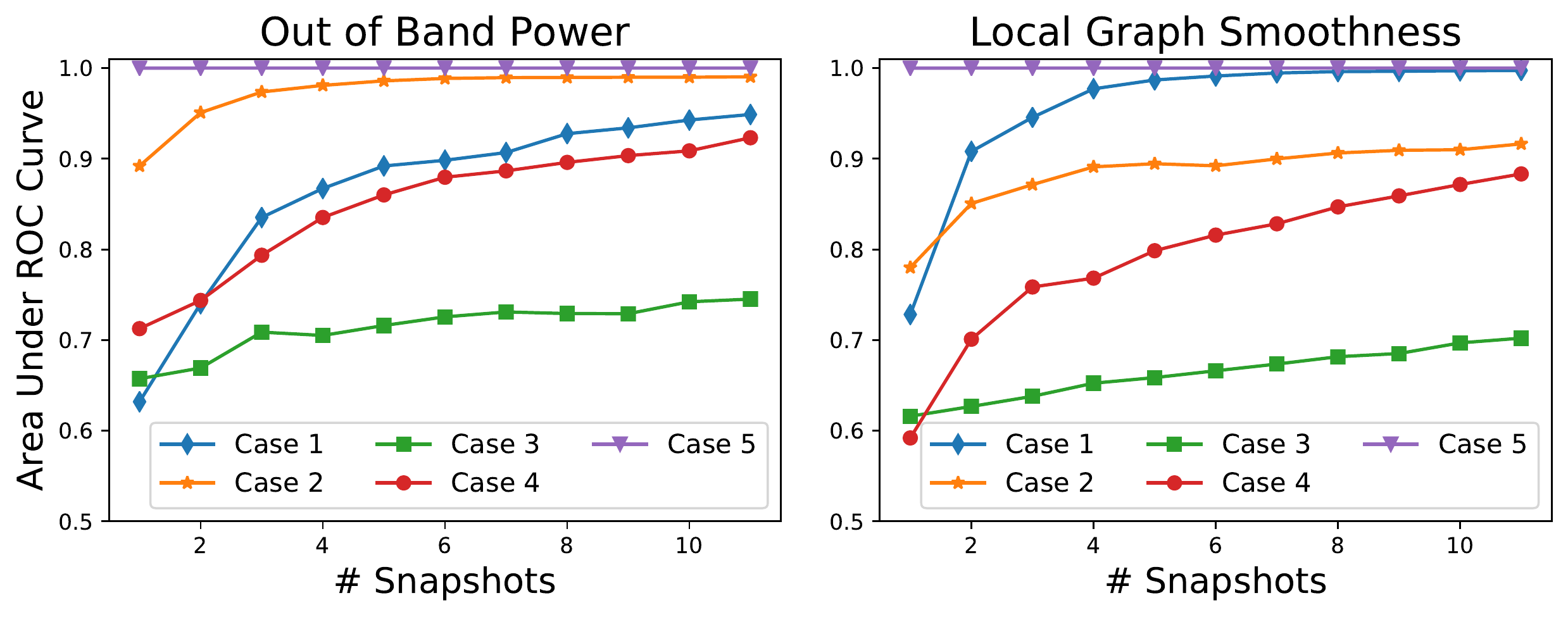}

    \caption{Area under the ROC curves for the five anomaly cases.} 
    \label{fig:couzin_auc_roc}

\end{figure}

Each of these model parameterizations was run with 100 Monte Carlo runs that varied the initial state of each agent and then iterated for 1500 time steps to allow for the swarm to stabilize.  After that, the graph was defined from agent states as described above and the two  approaches \ac{OOBP} and \ac{LGS} were evaluated at discrete time steps 50 time steps apart to create swarm ``snapshots'' to allow for correlations in swarm state to die out.  The results of the norm squared of the filtered signal at each agent were summed over the snapshots to produce the measurements that were thresholded to define our anomaly detector.  The area under the \ac{ROC} curve, empirically estimating the probability that an anomalous agent whose parameters are uniformly randomly selected from those explored would have a higher statistic value than a randomly chosen nominal agent, providing an indication of how effectively the threshold statistic can be used to detect anomalous agents \cite{hanley1982meaning}, is shown in Fig.~\ref{fig:couzin_auc_roc}.

\begin{figure}[htb]

    \centering
    \includegraphics[width=\columnwidth]{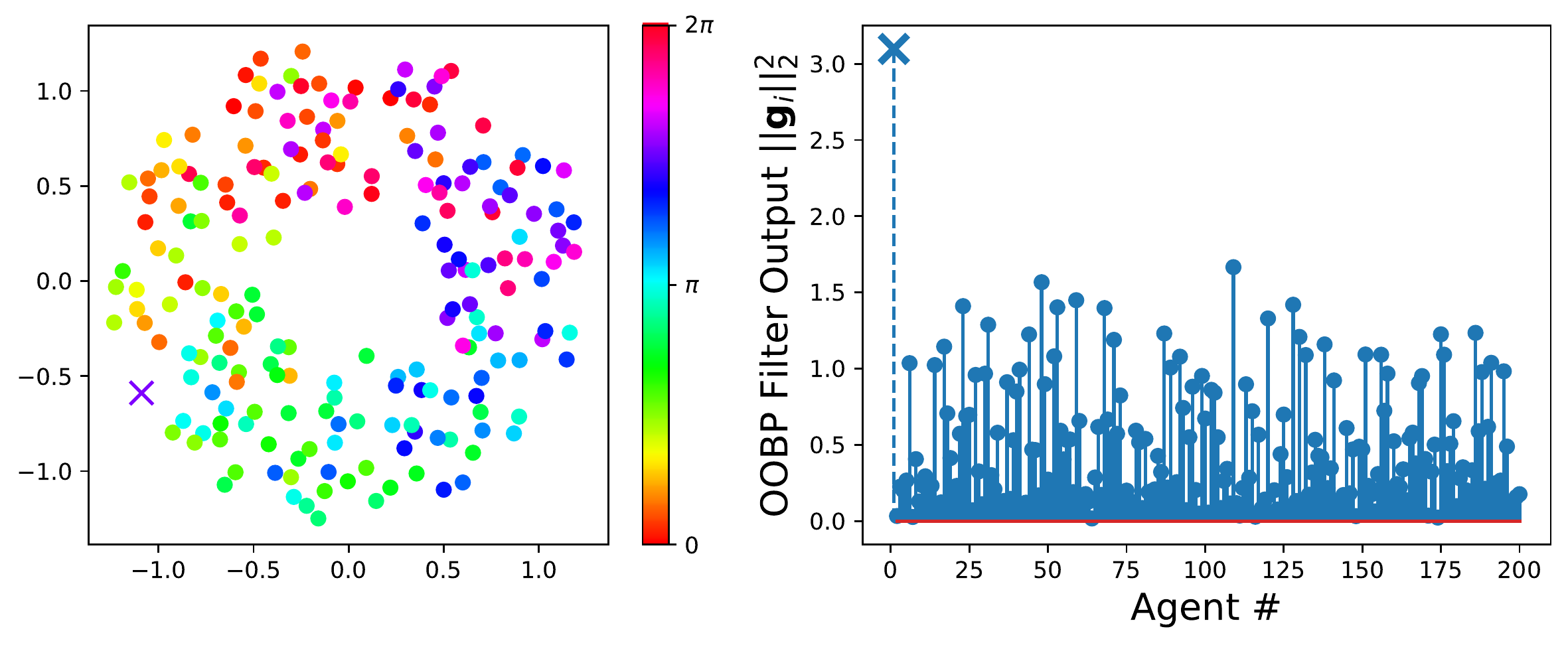}

    \caption{Swarmalator state with anomalous agent, marked by $\times$ in both panes. Left: Agent positions, with color indicating $\theta_j$. Right: Graph filtered response of $\mathbf{h}$.}
    \label{fig:swarmalator_anomalous}

\end{figure}

These curves show that there is a range of efficacy of both detectors across the cases considered, although we note that all of them produce areas substantially greater than 0.5 (random chance), and all appear to improve as the number of snapshots (analogous to integration time) increases.  It is clear that Case 5 appears to be an extremely easy detection problem, whereas Case 3 is the most challenging. Viewing a typical example from Case 5 (Fig.~\ref{fig:swarmalator_anomalous}) we see that the anomalous agent is somewhat out of phase from the rest of the swarm, but may be challenging to visually identify if not explicitly marked.  However, the filtered response is a clear outlier, consistent with the intuition behind our approach.
The more challenging cases have considerable overlap in the distributions of the filtered signals. We note that detecting the repulsion-based anomaly in Cases 1 and 2 appears to be considerably easier than detecting anomalous alignment behaviors in Cases 3 and 4. Anecdotally we do see that more repulsive agents will occasionally appear farther from $\bar{\mathbf{x}}$ than nominal agents, whereas the alignment effects are generally unnoticeable visually.
Another interesting facet of these results is that \ac{OOBP} appears to be a more effective detection mechanism, except in Case 1. Why this is the case despite such a concentrated \ac{GFT} response of the nominal behavior we leave to future research.

\section{Conclusion}
In summary, we demonstrated how the \ac{GFT} structure of swarms could be leveraged to detect anomalous agent behaviors in a number of scenarios.  We considered two different filter-based approaches and used a variety of graph signals as the subjects of the analysis.  Overall, we demonstrated that the relative efficacy of these approaches appears highly dependent on the specific context. Using detection-theoretic \ac{ROC} analysis we demonstrated that accumulating information over several snapshots of the swarming data may be needed to have effective detectors in the most challenging cases.


This work indicates a number of potential future directions in not only anomaly detection in swarms but more generally in the application of \ac{GSP} techniques. Further analysis of the cases considered here is warranted, and additional swarming models and anomalies could be analyzed as well. \ac{GSP} techniques that analyze both the vertex and time domains simultaneously could be incorporated \cite{verdoja2020graph}, although this would require handling time-varying graphs (rather than the independent manner in which they were treated here). In principle, this work could be combined with distributed graph filtering techniques to perform self-anomaly detection within the swarm.  Finally, we note that the the techniques here could be adapted to behavior discrimination and classification of agents in a swarm such as leader vs.\ follower behaviors.

\bibliographystyle{IEEEbib}
\bibliography{references}

\end{document}